\documentstyle[12pt]{article}
\def\V{{\cal V}}
\def\H{{\cal H}}

\newcommand{\be}{\begin{eqnarray}}
\newcommand{\en}{\end{eqnarray}}

\begin{document}
\begin{titlepage}
\begin{flushright}
EFI 95-45 \\
MPI-Ph/95-102  \\
\end{flushright}

\begin{center}
\vskip 0.3truein

{\bf {Superconvergence, Confinement and Duality}}
\footnote{Talk presented at the International Workshop on High Energy Physics,
Novy Svit, Crimea, September 1995. To be published in the Proceedings.}

\vskip0.5truein

{Reinhard Oehme}
\vskip0.2truein

{\it Enrico Fermi Institute and Department of Physics}

{\it University of Chicago, Chicago, Illinois, 60637, USA}
\footnote{Permanent Address}

{\it and}

{\it Max-Planck-Institut f\"{u}r Physik}

{\it - Werner-Heisenberg-Institut -}

{\it 80805 Munich, Germany}
\end{center}
\vskip0.2truein
\centerline{\bf Abstract}
\vskip0.13truein

Arguments for the confinement of transverse gauge field
excitations, which are based upon superconvergence relations
of the propagator, and upon the BRST algebra, are reviewed
and applied to supersymmetric
models. They are shown to be in agreement with recent results
obtained as a consequence of holomorphy and duality
in certain $N=1$ SUSY models.
The significance of the one loop anomalous dimension of the
gauge field in the Landau gauge is emphasized.
For the models considered, it is shown
to be proportional, with negative realtive sign,
to the one loop coefficient of the
renormalization group function for the {\it dual map} of the
original theory.

\end{titlepage}
\newpage
\baselineskip 20 pt
\pagestyle{plain}


\setcounter{equation}{0}

\vskip0.2truein

Recent advances in the understanding of the phase structure of
supersymmetric gauge theory models \cite{WIS,SEI,SEN}
make it possible to compare general arguments for the
confinement of transverse gauge field excitations \cite{ROC,NIC} with the
results obtained on the basis of holomorphy and duality.
For the specific supersymmetric models considered, we find
that our conclusions about confinement \cite{LOP},
which are related to the superconvergence
of the gauge field propagator, are in agreement with results
obtained on the basis of duality.
The comparison leads to
new insights concerning the physical significance of the
one-loop coefficient $\gamma_{00}$ of the gauge field
anomalous dimension in the Landau gauge ($\alpha = 0$, a fixed point in
$\alpha$).
Previous work on the structure and the asymptotic behavior of the
gauge field  propagator \cite{OWG,WZS,ROS}, and on confinement,
\cite{ROC,RLP} has already shown that the coefficient $\gamma_{00}$, and
in particular it's sign, can be of physical importance.

In this talk, I first briefly review the arguments, based upon
superconvergence,
for the confinement of transverse gauge field excitations in the presence of
a limited number of matter fields. Applying these methods to $N=1$
supersymmetric gauge theories in the Wess-Zumino gauge, I describe how the
appropriate coefficient $\gamma_{00}$ for the gauge field becomes proportional
to the negative
of the one loop coefficient $\beta^d _0 $ of the renormalization
group function $\beta^d (g^2)$ for the {\it dual map} of the
gauge theory considered.
This feature is of direct relevance for confinement. The ``electric'' and
the ``magnetic'' versions of the theory should have the same low energy
properties. Since $\gamma_{00}<0$ is associated with $\beta^d_0>0 $,
it implies IR-freedom for the magnetic formulation, which is then the
appropriate low energy description in terms of composites formed by the
confined, elementary, electric quanta. The comparisons I discuss here
are very preliminary. Much work remains to be done in this area. In view
of the specific assumptions made in applying superconvergence arguments,
there may be many models where a simple comparison is not possible.

\vskip0.2truein
In previous work \cite{ROC}, we have considered confinement on the basis
of analytic and asymptotic properties of the gauge field
propagator within the framework of a covariant formulation
of the gauge theory \cite{BRS}. Using the BRST cohomology, and the
assumption of completeness of the BRST operator \cite {SPI, ROC},
a covariant physical space $\H$ is defined within the general
state space $\V$ of indefinite metric.
Confined  excitations are then states, which are not elements of
$\H$. With other unphysical quanta, like ghosts and longitudinal and
time-like gauge excitations, they form quartet representations
of the BRST algebra in $\V$ \cite{KOJ}.

Let $D(k^2)$ be the structure function for the transverse gauge field
propagator. It follows from Lorentz-covariance and the spectral condition,
that this function is analytic in the cut $k^2$-plane with branch lines
along the positive real axis. Using renormalization group methods,
we find for all linear, covariant
gauges ($\alpha \geq 0 $) , and for $ k^2 \rightarrow \infty $
in {\it all directions} \cite{OWG} of the complex $k^2$-plane:
\begin{eqnarray}
-k^2 D (k^2,\kappa^2,g,\alpha)&\simeq& \frac{\alpha}{\alpha_0}
 + C_R (g^2, \alpha) \left(-\beta_0 \ln
\frac{k^2}{\kappa^2}\right)^{-\gamma_{00}/\beta_0} + \cdots ~.
\label{1}
\end{eqnarray}
The corresponding asymptotic terms for the discontinuity along the
positive, real $k^2$--axis are given by
\begin{eqnarray}
-k^2 \rho (k^2,\kappa^2,g,\alpha) & \simeq& \frac{\gamma_{0 0}}{\beta_0}
C_R (g^2,\alpha) \left(-\beta_0 \ln \frac{k^2}{\vert
\kappa^2\vert}\right)^{-\gamma_{0 0}/\beta_0 - 1} + \cdots ~.
\label{2}
\end{eqnarray}
Here $\kappa^2 < 0 $ is the normalization point, and
\be
\gamma(g^2,\alpha) &=& (\gamma_{00} + \alpha \gamma_{01} ) g^2 ~+~
\cdots ~, \cr
\beta(g^2) &=& \beta_0 g^4 ~+~ \cdots ~
\label{2a}
\en
are the limits
$g^2 \rightarrow 0$ of the anomalous dimension and the
renormalization group function, while $\alpha_0 = - {\gamma_{00} }/
{\gamma_{01}}$. It is always assumed that we have asymptotic
freedom, so that $\beta < 0$.

We reproduce the formulae given above, in order to
emphasize the importance of the one loop coefficient
$\gamma_{00} = \gamma_0 (\alpha = 0)$ in determining the essential
asymptotic term in {\it all} gauges. In the derivation
of these asymptotic expressions, we have made
the assumption that the exact propagator connects with the
perturbative expression in the weak coupling limit $g^2 \rightarrow +0$,
at least as far as the leading term is concerned.

In this talk, we consider massless gauge theories, but intrinsic
masses may be accommodated by the use of mass independent
renormalization schemes \cite{WEI}. We also use the Landau gauge. The use
of other covariant gauges is possible, but more complicated.
Since confinement is a physical notion, it is sufficient to argue
in a particular gauge.

It follows from Eq.(\ref{1}) for $\alpha = 0$ ,
that $D(k^2)$ vanishes faster than $k^{-2}$ for
$k^2 \rightarrow \infty $ in all directions, provided we have
$\gamma_{00} < 0~,~\beta_0 < 0 $. Consequently,
we get the superconvergence relation \cite{WZS,ROS}
\be
\int_{-0}^{\infty}d k^2 \rho (k^2, \kappa^2, g )
{}~~= ~~0 ~.
\label{3}
\en
On the other hand, there is {\it no} superconvergence
for $\gamma_{00} > 0~,~\beta_0 < 0 $. The superconvergence relation
(\ref{3}) gives a direct connection between high and low energy
properties of the gauge theory. We note that the discontinuity
$\rho$ represents the norm of the states ${\tilde{A}}^{\mu\nu} (k)
\vert 0 \rangle $, where $ A^{\mu\nu} = \partial^{\mu} A^{\nu} -
\partial^{\nu} A^{\nu} $. We have
\begin{eqnarray}
&~&\langle 0 | {\tilde{A}}^{\mu \nu}_{a}(k')
{\tilde{A}}^{\varrho\sigma}_{b}(-k) |0\rangle~~ =~~
 \delta_{ab} \theta (k^0) \delta (k'-k) \pi \rho (k^2) \cr
&~& \times (-2){(2\pi)}^4 \left( k^\mu k^\varrho g^{\nu \sigma} -
k^\mu k^\sigma g^{\nu \varrho}
+ k^\nu k^\sigma g^{\mu\varrho} - k^\nu k^\varrho g^{\mu\sigma}\right) ~~.
\label{3a}
\end{eqnarray}
With test functions $C^a_{\mu\nu} (k)$, we form states
\be
\Psi (C) ~= ~ \int d^4 k C^a_{\mu\nu} (k) {\tilde{A}}^{\mu\nu}_a (-k)|0\rangle
{}~,
\label{3aa}
\en
and  obtain the norm
\begin{eqnarray}
\left( \Psi (C), \Psi (C) \right)~ &=&~ \int d^4 k \theta(k^0)
\pi \rho (k^2) C(k) ~~, \cr
C(k)~ &=&~ -8(2\pi)^4 k^{\mu}{\overline{C}}^a_{\mu\nu}(k) k^{\rho}
C^a_{\rho\sigma} g^{\nu\sigma} ~~,
\label{3ab}
\end{eqnarray}
where $C(k) > 0$  for $k^2 \geq 0, ~k^0 \geq 0$.

Our arguments for confinement make intensive use of renormalization
group methods and of the BRST algebra. We refer to Ref.\cite{ROC}
for a discussion of the many details. In essence, we consider the
functions $D_{+}(k^2)$ and $D_p (k^2)$, which are given by representations
of the form
\be
D_{+,p}(k^2)~~ =~~ \int d{k'}^2 \frac{\rho_{+,p} ({k'}^2)}{{k'}^2-k^2} ~~,
\label{3b}
\en
where $\rho_{+}(k^2) $ is the contribution to $\rho(k^2)$ from all positive
norm states of the form ${\tilde {A}}^{\mu\nu}|0\rangle $,
while $\rho_p (k^2)$ is due to representatives of the corresponding
physical states. We show that these separations are exclusive, covariant, and
defined a fashion which is invariant under renormalization group
transformations.
For both functions, the renormalization group gives a connection between the
behavior for $k^2 \rightarrow \infty$, $g^2$ fixed and $g^2 \rightarrow 0$,
$k^2$ fixed. If $\rho_p \neq 0$, and there are physical gauge field quanta
(gluons in QCD), the weak coupling limit $g^2 \rightarrow 0$ of $D_p$ and
$\rho_p$ should be bounded and related to the perturbative gluon pole term.
But for $\gamma_{00}<0,~\beta_0<0$ , it follows from our analysis that
$D_p$ and $\rho_p$ diverge for $g^2 \rightarrow 0$ at least like
$(g^2)^{-\gamma_{00}/\beta_0}$. Hence we are led to the conclusion that
$\rho_p = 0$ in this region of $N_F$, and all positive norm states of the
form ${\tilde{A}}^{\mu\nu} |0 \rangle $ are not elements of the physical
state space $\H$. The alternative to the divergence of $D_p$ would be
superconvergence for $k^2 \rightarrow \infty$, which also implies $\rho_p = 0$,
because of the positive norm. We refer to Ref.\cite{ROC} for questions about
multipole terms and other details.

In contrast to the situation described above, there is no superconvergence
for the region of $N_F$ where $\gamma_{00}>0, ~\beta_0<0 $, and no divergence
of $Dp$ in the weak coupling limit. Our argument for confinement is not
applicable, although there could be other reasons. On the other hand,
in this range, a theory like QCD looks quite compatible with
the existence of transverse gluons in the physical space $\H$.
It is therefore possible \cite{ROC,NIC}, that there is a
phase transition as $N_F$ increases through the point where $\gamma_{00}(N_F)$
has a zero.
For QCD, this zero occurs between $N_F=9$ and $N_F=10$, so that our argument
implies confinement for $N_F \leq 9$. More general, for a gauge theory
with $G=SU(N_C)$, and with matter fields in the regular representation,
we have the zero of $\gamma_{00}$ at $N_F=\frac{13}{4} N_C $, and that of
$\beta_0$ at $N_F=\frac{22}{4}N_C$. We get confinement for
$N_F<\frac{13}{4}N_C$,
and there is the possibility of unconfined gauge quanta in the interval
$\frac{13}{4}N_C < N_F < \frac{22}{4}N_C$.

The general conclusions concerning confinement, in the sense of the
absence of excitations from the physical state space $\H$ for
$\gamma_{00} < 0 , ~~\beta_0 < 0 $, are supported by more heuristic
considerations. Let us discuss the potential between static color
charges, to use the language of QCD. The quark-antiquark potential
is related to the structure function $D(k^2)$ in a well known fashion
\cite{WIL}.
Using the Landau gauge, we see from Eq. (\ref{1}) that $D(k^2)$ has
an asymptotic behavior compatible with a
dipole representation \cite{RLP,NIP}
\begin{eqnarray}
D(k^2)~ &=&~ \int_{-0}^{\infty} d {k'}^2
 \frac{\sigma ({k'}^2)}
{({k'}^2 - k^2)^2} ~~, \cr
\sigma (k^2)~ &=&~ \int_{-0}^{k^2} d {k'}^2 \rho({k'}^2) ~~.
\label{3c}
\end{eqnarray}
The asymptotic limit of $\sigma (k^2)$ is given by
\be
\sigma (k^2) ~\simeq C_R (g^2,0) \left(lg (k^2 / |\kappa^2|)
\right)^{-\gamma_{00}/\beta_0} ~,
\label{3d}
\en
with the positive coefficient $C_R (g^2, 0) > 0$, where
\be
C_R(g^2, 0)~ = ~ exp \{ \int_{g^2}^0 dx \left(\frac{\gamma(x. 0)}{\beta(x)}~ -
{}~ \frac{\gamma_{00}}{\beta_0 x} \right) \} ~,
\label{3e}
\en

We first consider the case $\gamma_{00} < 0 $. Using the relations given
above, we find, for large values of $k^2 > K^2$,
\be
\sigma (\infty ) = 0~, ~~\sigma (k^2) > 0~, ~~{\sigma}' (k^2) = \rho (k^2) < 0
{}~.
\label{3f}
\en
Here $K^2 \geq 0 $ is the largest finite zero of $\rho (k^2)$. Since $\rho$ is
negative asymptotically, and since it cannot be negative for all $k^2 \geq 0 $
in view of the superconvergence relation (\ref{3}), there must be at least one
zero. The position $K^2 $ is renormalization group invariant: $K^2(g^2,
{\kappa}^2)
= K^2({g'}^2, {\kappa'}^2) $, with $g' = \overline{g} ({\kappa'}^2 /\kappa^2 ,
g)$,
and it is easily shown to vanish exponentially for $g^2 \rightarrow +0 $
proportional to $ exp(\frac{1}{\beta_0 g^2}) $. The dipole weight function
$\sigma$ has a maximum at $K^2$, from where it decreases to zero as
$k^2 \rightarrow \infty $. We do not know it's shape for $k^2$ below the
maximum, except that it is identically zero for negative values of $k^2$.
If the maximum at $K^2$ is dominant, we may crudely approximate $\sigma$ by a
delta function, so that the structure function $D(k^2)$ has the form
$(k^2 - K^2)^{-2} $ , corresponding to an approximately linear potential
for $R \ll 1/\sqrt {K^2 } $. For a general discussion, also covering the
question of possible multipole contributions to $\rho $, may be found in
\cite{RLP}.
Multipole terms do not contribute to the superconvergence relation (\ref{3}).

For values of $N_F$ where $\gamma_{00} > 0, ~\beta_0 < 0 $, we have
for $k^2 > K^2$, in contrast
to Eq.(\ref{3f}),
\be
\sigma (\infty ) = \infty~, ~~\sigma (k^2) > 0~, ~~{\sigma}' (k^2) = \rho (k^2)
> 0 ~.
\label{3g}
\en
There is no superconvergence relation, and $\sigma $ has a minimum at $K^2$.
We have no indication of an approximately linear potential. Of course, from a
mathematical point of view, we cannot exclude the possibility that a dipole
term is present in the structure function $D(k^2)$ at $k^2 = 0$,
giving a linear potential independent of the sign of $\gamma_{00}$.
Nevertheless,
if we rely only on the ordinary spectrum of states. the potential
considerations
support the possibility of a de-confinement transition as the coefficient
$\gamma_{00} (N_F) $ changes sign with increasing $N_F$.

There are no superconvergence relations for the quark propagator.
However, one can connect the results described above with
a criterion for general color confinement given by Kugo and
Ojima \cite{KOJ}, and discussed further by Nishijima
\cite{NLP}.

\vskip0.2truein

In Ref.\cite {LOP}, we have applied our arguments for gluon
confinement to supersymmetric theories, concentrating mainly
on the gauge field propagator in the Wess-Zumino gauge.
One obtains then conditions for the transverse gauge
field excitations, which are elementary quanta in the formulation
of the theory, to be excluded from the physical state space
$\H$.

For $N = 1$ supersymmetric gauge theories, we write the one-loop
coefficients of the function $\beta (g^2)$ and the anomalous dimension
$\gamma (g^2, \alpha = 0)$ in the form
\be
\beta_0 ~~=~~ (16\pi^2)^{-1} \left( -3 C_2 (G) ~~+~~\sum_{i} n_i
T(R_i) \right) ~~,
\label{4}
\en
and
\be
\gamma_{00} ~~=~~ (16\pi^2)^{-1} \left( -\frac{3}{2} C_2 (G) ~~+~~
\sum_{i} n_i T(R_i) \right) ~~,
\label{5}
\en
where $n_i$ is the number of $N = 1$ chiral superfields in the
representation $R_i$. These coefficients are determined by
the elementary field content of the theory. We have
\be
\gamma_{00} ~~<~~ 0~~, ~~\beta_0 ~~<~~ 0 ~~~ for \cr
\sum_{i} n_i T(R_i) ~~ <~~ \frac{3}{2} ~ C_2 (G) ~~,
\label{6}
\en
which is the condition for the validity of the supercovergence
relation (\ref{3}) for the structure function, and hence for
the confinement of the elementary transverse gauge field excitations.
As an example, for $G = SU(N_C)$,
and matter fields in the fundamental representation
$N_F \times ({\bf N_C} + {\bf {\overline {N}}_C})$,
we have $\gamma_{00} < 0 $,
$\beta_0 < 0 $ for $N_F < \frac{3}{2} N_C $, where $N_F$ refers to
four-component spinors in contrast to $n_i$. As mentioned before,
the superconvergence argument implies confinement for
$N_F < \frac{3}{2} N_C $, and we have indications of a de-confining
phase transition at $N_F = \frac{3}{2} N_C$ , as $N_F$ increases
\cite{LOP}.

Let us now compare these results with the picture obtained on the
basis of electric-magnetic duality for $N = 1$ supersymmetric
gauge theories. In particular, we are interested in the r\^ole
of the coefficient $\gamma_{00} = \gamma_0 (\alpha = 0) $ in the
expansion of the gauge field anomalous dimension, which was
important in the superconvergence arguments. It turns out to
be of a more universal significance than expected, a priori,.
comparable to the one loop coefficient $\beta_0$ of the renormalization
group function. Using the model described above with $G = SU(N_C)$,
it is argued that, for appropriate values of $N_C$ and $N_F$, the
electric gauge theory has a dual magnetic description (dual map)
given by the gauge theory with $G^d = SU(N_C - N_F)$, where $N^d_F =
N_F$ and $N^d_C = N_F - N_C$, plus a number of colorless, massless
meson fields \cite{SEN}. This dual map is
introduced in order to have a solution of the 't Hooft anomaly
matching conditions for $N_F > N_C + 1 $. It describes massless
magnetic excitations, which are composites of the elementary
electric states, and which are present in addition to invariant
bound states.

The one loop coefficients of both theories are given by:
\be
G = SU(N_C) ~~~"electric"~~~ N = 1 ~~ SUSY \cr
{}~~~~~~~~~~~~~~~~~~~~~~~~~~~~~~~~~~~~~~~~~~~~\cr
\beta_0 ~~~=~~~(16\pi^2)^{-1} (-3N_C ~+~ N_F ) \cr
\gamma_{00} ~~~=~~~(16\pi^2)^{-1} (-\frac{3}{2} N_C ~+~ N_F) ~~,
\label{7}
\en
and
\be
G = SU(N_F-N_C) ~~~"magnetic"~~~N = 1 ~~ SUSY  \cr
{}~~~~~~~~~~~~~~~~~~~~~~~~~~~~~~~~~~~~~~~~~~~~~~~\cr
\beta^d_{0} ~~~=~~~(16\pi^2)^{-1} (-2N_F ~+~ 3N_C)  \cr
\gamma^d_{00} ~~~=~~~(16\pi^2)^{-1} (-\frac{1}{2} N_F ~+~ \frac{3}{2}
N_C ) ~.
\label{8}
\en
Here the coefficients of the dual map have been evaluated at the
same number of flavors $N^d_F = N_F $ as the original, electric
theory. However, these flavors refer to representations of the
magnetic gauge group.

{}From the above equations, we can extract the following important
relationships between electric and magnetic coefficients:
\be
\beta_0 (N_F) ~&=&~ - 2 \gamma^d_{00} (N_F) ~~,
\label{9}
\en
\be
\beta^d_0 (N_F) ~&=&~ - 2 \gamma_{00} (N_F) ~~,
\label{10}
\en
where it is again understood, that the variable $N_F$ on both sides
refers to matter fields with different quantum numbers in the electric
and magnetic functions respectively.

These duality relationships are not restricted to the particular
model considered. For example, $N=1$ supersymmetric
gauge theory with the group $G = SO(N_C)$ and with $N_F$
flavors in the representation $N_F \times {\bf N_C}$, has a dual map
with the group $G^d = SO(N_F-N_C+4)$ \cite{IOS}. The duality relations are
again given by equations (\ref{9}) and (\ref{10}). For this supersymmetric
theory with $G = SO(N_C)$, the coefficient $\gamma_{00}(N_F)$ changes
sign at $N_F = \frac{3}{2} (N_C - 2)$. It is certainly of
interest to study further models.

At least within the framework of the $N=1$ supersymmetric models
discussed here, the relations (\ref{9}) and (\ref{10})
underline our previous
results concerning the significance of the anomalous dimension
coefficient, and the phase transition associated
with its change of sign.
Let us assume, at first, that $N_C$ and $N_F$ are
sufficiently large, so that there is a non-trivial dual map.
For the $SU(N_C)$ model in the region $N_F < \frac{3}{2} N_C $,
the transverse gauge quanta (gluons) are not in the physical state
space $\H$. In the dual magnetic picture, the $\beta$-function
coefficient $\beta^d_0 (N_F) $ is positive for $N_F < \frac{3}{2}
N_C$, so that we have infrared freedom for the magnetic theory,
in contrast to the asymptotic freedom of
 the the theory in the electric version.
Therefore, in this region, the low energy description should be
in terms of the magnetic excitations, which may be viewed as
composites of the elementary electric quanta. The latter are not
in the physical space $\H$, in agreement with the superconvergence
result. The nature of the physical composites cannot be obtained
from the superconvergence argument, but requires information about
the character of the dual map.

There is a non-Abelian dual map of the $SU(N_C)$ supersymmetric
theory provided $N_F \geq N_C + 2 $. The point $N_F = N_C + 2$
is below the critical number
$N_F = \frac{3}{2} N_C $ only for $N_C \geq 4 $.
If applicable, we have $\beta^d_0 (N_F) > 0 $ in the range
$ N_C + 2 < N_F < \frac{3}{2} N_C $, and the low energy system is
described by massless magnetic composites as discussed before.
There are no electric elementary quanta in the physical space
$\H$. For even smaller values of $N_F$,
like $N_F = N_C + 1$ and $N_F = N_C $,
the Higgs mechanism has removed the massless quanta of the dual
gauge theory. We have massive composites as physical states in
$\H$.

Let us now discuss the interval $\frac{3}{2}N_C < N_F < 3N_C $, where we have
$\gamma_{00} > 0 $ and $\beta_0 < 0 $ for the electric theory. There is
asymptotic freedom, but no supercovergence of the structure function. The
corresponding argument for the confinement of transverse gauge quanta,
in the sense of their absence from the physical state space $\H$, is not
applicable. As we have mentioned, there are
indications for non-confinement, so that there could
be a phase transition at $N_F = \frac{3}{2}N_C $.

If we consider again the dual magnetic formulation with the gauge group
$G^d = SU(N_F - N_C)$, where we have written $N^d_F = N_F $, the interval
$\frac{3}{2}N_C < N_F < 3N_C $ in the electric case corresponds to
$ 3(N_F - N_C) > N_F > \frac{3}{2}(N_F - N_C) $ for the magnetic theory.
In this interval, we have $\gamma^d_{00}(N_F) > 0 $ and $\beta^d_0 (N_F)
< 0$, so that both theories have asymptotic freedom and no superconvergence,
but with different gauge groups and different matter contents.

At large distances, it has been argued \cite{SEI,SEN}
that the electric theory has an IR fixed point, so that we have an
interacting, conformal field theory with unconfined excitations
and an $R^{-1}$ potential between static charges. This is in
agreement with the possibility of a de-confining phase transition at
$N_F = \frac{3}{2}N_C $, which we have discussed on the basis of
superconvergence.
Corresponding arguments apply to the magnetic theory, which would the also be
conformally invariant at low energies. The problem of duality of these
conformal theories has been discussed in \cite{SEN,KSV}.

\vskip0.2truein

For the supersymmetric models considered, criteria for the confinement
of transverse gauge particles, which are based upon superconvergence
of the propagator, are in agreement with the results of the
duality approach, but other models should be studied.
We recall that the superconvergence arguments are also
applicable to non-supersymmetric gauge theories like QCD.

\vskip0.7truein
\centerline{\bf ACKNOWLEDGMENTS}
\vskip0.2truein

I am grateful to K. Sibold for helpful conversations.
It is a pleasure
thank Wolfhart Zimmermann, and the
Theory Group of the Max Planck Institut
f\"{u}r Physik, Werner Heisenberg Institut, for their
kind hospitality in M\"{u}nchen.

This work has been supported in part by the
National Science Foundation, grant PHY 91-23780.

\newpage
\vskip0.7truein

\end{document}